\documentclass[twoside,slac_one]{revtex4}
\usepackage{graphicx}
\usepackage{fancyhdr}
\usepackage{amsmath} 
\usepackage{bm}
\usepackage{amsxtra}
\usepackage{amssymb}
\usepackage{amsthm}
\usepackage{wasysym}
\usepackage{latexsym}
\usepackage{lscape}
\usepackage{epstopdf}
\usepackage[usenames,dvipsnames]{color}
\usepackage{color}

\newcommand{\sub}[1]{\mathrm{\scriptscriptstyle{#1}}}

\newcommand{\magHg}{\ensuremath{{}^{199}\mathrm{Hg}}}

\newcommand{\parafield}{\ensuremath{\uparrow\!\uparrow}}
\newcommand{\aparafield}{\ensuremath{\uparrow\!\downarrow}}

\newcommand{\nedm}{\ensuremath{d_{\sub{n}}}}

\newcommand{\ecm}{\ensuremath{e\!\cdot\!\mathrm{cm}}}

\newcommand{\fTHz}{\ensuremath{\mathrm{fT/\sqrt{Hz}}}}
\newcommand{\pTHz}{\ensuremath{\mathrm{pT/\sqrt{Hz}}}}

\begin{document}

\title{An Improved Search for the Neutron Electric Dipole Moment}

%
\author{M. Burghoff}
\author{A. Schnabel}
\affiliation{Physikalische Technische Bundesanstalt, Berlin,
Deutschland}
\author{G. Ban}
\author{T. Lefort}
\author{Y. Lemiere}
\author{O. Naviliat-Cuncic}
\author{E. Pierre\footnotemark[3]}
\author{G. Qu\'{e}m\'{e}ner}
\affiliation{LPC Caen, ENSICAEN, Universit\'{e} de Caen, F-14050
Caen, France}
\author{J. Zejma}
\affiliation{Jagellonian University, Cracow, Poland}
\author{M. Kasprzak}
\author{P. Knowles}
\author{A. Weis}
\affiliation{University of Fribourg, CH-1700, Switzerland}
\author{G. Pignol}
\author{D. Rebreyend}
\affiliation{LPSC, Universit\'{e} Joseph Fourier Grenoble 1,
CNRS/IN2P3, INPG, F-38026 Grenoble, France }
\author{S. Afach}
\author{G. Bison}
\affiliation{Biomagnetisches Zentrum, Jena, Germany}

\author{J. Becker}
\author{N. Severijns}
\author{S. Roccia\footnotemark[4]}
\affiliation{Instituut voor Kern - en Stralingsfysica Katholieke
Universiteit Leuven, B-3001 Leuven, Belgium}
\author{C. Plonka-Spehr}
\author{J. Zenner\footnotemark[3]}
\affiliation{Institut f\"ur Kernchemie,
Johannes-Gutenberg-Universit\"at, D-55128 Mainz, Germany}
\author{W. Heil}
\author{H.~C. Koch\footnotemark[5]}
\author{A. Kraft}
\author{T. Lauer}
\author{Yu. Sobolev\footnotemark[6]}
\affiliation{Institut f\"ur Kernchemie,
Johannes-Gutenberg-Universit\"at, D-55128 Mainz, Germany}
\author{Z. Chowdhuri}
\author{J. Krempel}
\author{B. Lauss}
\author{A. Mtchedlishvili}
\author{\underline{P. Schmidt-Wellenburg}\footnotemark[1]}
\author{G. Zsigmond}
\affiliation{Paul Scherrer Institut (PSI), CH-5232 Villigen-PSI,
Switzerland}
\author{M. Fertl\footnotemark[3]}
\author{B. Franke\footnotemark[3]}
\author{M. Horras\footnotemark[3]}
\author{K. Kirch}
\author{F. Piegsa}
 \affiliation{ETH Z\"urich, CH-8093
Z\"urich, Switzerland}

\begin{abstract}
A permanent electric dipole moment of fundamental spin-$1/2$
particles violates both parity~({\it P}\,) and time reversal~({\it
T}\,) symmetry, and hence, also charge-parity~({\it CP}\,) symmetry
since there is no sign of {\it CPT}-violation. The search for a
neutron electric dipole moment~(nEDM) probes CP violation within and
beyond the Standard Model. The experiment, set up at the Paul
Scherrer Institute~(PSI), an improved, upgraded version of the
apparatus which provided the current best experimental limit,
$d_{\sub{n}}< 2.9\times10^{-26}\,\ecm$~(90\,\% C.L.), by the
RAL/Sussex/ILL collaboration: {\it Baker et~al., Phys. Rev.
Lett.~97, 131801~(2006)}. In the next two years we aim to improve
the sensitivity of the apparatus to
$\sigma(\nedm)=2.6\times10^{-27}\,\ecm$ corresponding to an upper
limit of $d_{\sub{n}}< 5\times10^{-27}\,\ecm$~(95\,\% C.L.), in case
for a null result. In parallel the collaboration works on the design
of a new apparatus to further increase the sensitivity to
$\sigma(d_{\sub{n}}) = 2.6\times10^{-28}\,\ecm$.
\end{abstract}

\maketitle

\renewcommand{\thefootnote}{\fnsymbol{footnote}}

\footnotetext[1]{email: philipp.schmidt-wellenburg@psi.ch}
\footnotetext[3]{also Paul Scherrer Institut (PSI), CH-5232
Villigen-PSI, Switzerland} \footnotetext[4]{now at CSNSM,
CNRS-IN2P3, Universit\'{e} Paris-Sud, Orsay, France}
\footnotetext[5]{also University of Fribourg, CH-1700, Switzerland}
\footnotetext[6]{also PNPI, Gatchina, Russia}

\thispagestyle{fancy}


\section{\label{Introduction}Introduction}
The coexistence of a permanent electric, and magnetic dipole moment
of the neutron, intrinsically violates both {\it T} and {\it P}
symmetry, and assuming {\it CPT} conservation also {\it CP}\,. The
Standard Model~(SM) of particle physics gives a satisfactory account
for {\it P}-violation in weak interactions mediated by charged
($\mathrm{W^\pm}$) and neutral ($\mathrm{Z^0}$) weak currents as
well as for {\it CP}-violation in the decay of K- and B-mesons.
However, this source of {\it CP}-violation is far too weak to
explain the observed baryon asymmetry of the universe. The discovery
of a neutron electric dipole moment could thus help to explain and
may even solve this discrepancy. The electroweak SM predicts a
neutron electric dipole moment (nEDM) at $\nedm =
10^{-32\pm1}\,\ecm$. This sensitivity is beyond current experimental
techniques, hence any experimental sign of an nEDM will also be a
sign of physics beyond the Standard Model.
\newline
Our collaboration pursues this quest in two overlapping phases:

\begin{itemize}
        \item In a first step we will measure with an upgraded version of the
        original RAL/Sussex/ILL spectrometer. This apparatus is
        using UCN in vacuum at room temperature and
        delivered the last and most stringent limit so far on the
        nEDM of $\nedm < 2.9\times 10^{-26} \ecm$, C.L.~90\,\%\,\cite{Baker2006}.
        This result was statistically limited by the available UCN
        density at the Institut Laue Langevin~(ILL). The new UCN source at Paul Scherrer
        Institute~(PSI), now ramping up, will provide $\sim 25$
        times higher UCN densities. Together with an improved control on systematic
         effects using both, localized and large area magnetometers will
result in a sensitivity of $\sigma(\nedm)=2.6\times 10^{-27}\ecm$.
        \item In parallel, an entirely new apparatus (n2edm) is being designed. It
        will make optimal use of the high UCN density of the PSI
        source, and will have an improved active and passive magnetic
        shielding. This will improve the upper limit to $\nedm<5\times
        10^{-28}\,\ecm$, C.L.\,95\,\% in the case of a null result.

\end{itemize}

\section{Experimental Method}
The electric dipole moment $d_{\sub{n}}$ of the neutron is measured
by comparing the neutron precession frequency $\omega_{\sub{n}}$ in
parallel ($\uparrow \uparrow$) and anti parallel ($\uparrow
\downarrow$) electric $E$ and magnetic field $B$ configurations:

\begin{equation}
    \hbar\omega_{\sub{n}, \uparrow\!\uparrow\,/\,\uparrow\!\downarrow}=2\cdot \left|\mu_{\sub{n}} B\pm d_{\sub{n}}E\right|,
\end{equation}

\noindent where $\mu_{\sub{n}}$ is the magnetic moment of the
neutron. Taking the difference of both measurements allows us to
deduce the electric dipole moment:

\begin{equation}
    d_{\sub{n}}=\frac{\hbar\Delta\omega}{2\left(E_{\sub{\parafield}}+E_{\sub{\aparafield}}\right)}+
    \frac{\mu_{\sub{n}}\left(B_{\sub{\parafield}}-B_{\sub{\aparafield}}\right)}{\left(E_{\sub{\parafield}}+E_{\sub{\aparafield}}\right)}.
    \label{Eq:neutronEDM}
\end{equation}

The statistical sensitivity for measuring \nedm{} is then given by:

\begin{equation}
    \sigma\left(\nedm\right)=\frac{\hbar}{2\alpha T E \sqrt{N}},
\end{equation}
\noindent where $N$ is the number of neutrons, $E$ the strength of
the electric field, and $\alpha=\alpha_0e^{-\Gamma T}$ the neutron
polarization at the end of the cycle. The polarization depends on
$\alpha$, the product of initial polarization and analyzing power,
and the transversal depolarization rate $\Gamma$. Increasing the UCN
density by a factor 25 with the new UCN source and slightly
improving $\alpha$, $T$, and $E$ gives the goal sensitivity of
$\sigma(\nedm) = 2.6\times 10^{-27}\ecm$ in 400 nights.

However, the sensitivity of the measurement is dominated by the
second term of Eq.\,\ref{Eq:neutronEDM}, the stability of the
magnetic field in both configurations. Obviously, a very sensitive
measurement of the magnetic field in the neutron precession region
would allow to correct for the second term. Such a cohabiting large
area magnetometer was first implemented in~1997 by the
RAL/Sussex/ILL collaboration in the current experimental
apparatus\,\cite{Baker2006}. It uses polarized \magHg{} atoms
precessing within the neutron precession chamber, for a description
in more detail see Ref.\,\cite{Green1998}.

However, severe systematic effects arising from magnetic field
gradients within the precession chamber cannot be compensated for by
a co-magnetometer. Due to their small kinetic energy, the centers of
mass of UCN and of mercury atoms is separated by $\Delta h =
2-3\,\mathrm{mm}$. In a perfectly homogeneous magnetic field this
would still make no difference, however in vertical magnetic field
gradient the mercury atoms and the UCN will sense and average
differently the magnetic field. Therefore, an array of 12~Cs
optically pumped magnetometers~(Cs-OPM) have been adapted to the
special needs of the experimental apparatus, including vacuum and
high-voltage compatibility\,\cite{Knowles2009}. It will give us an
extra handle on systematic effects correlated to vertical magnetic
field gradients.

\section{The RAL/Sussex/ILL apparatus at PSI}
\begin{figure*}[t]
\centering
\includegraphics[width=135mm]{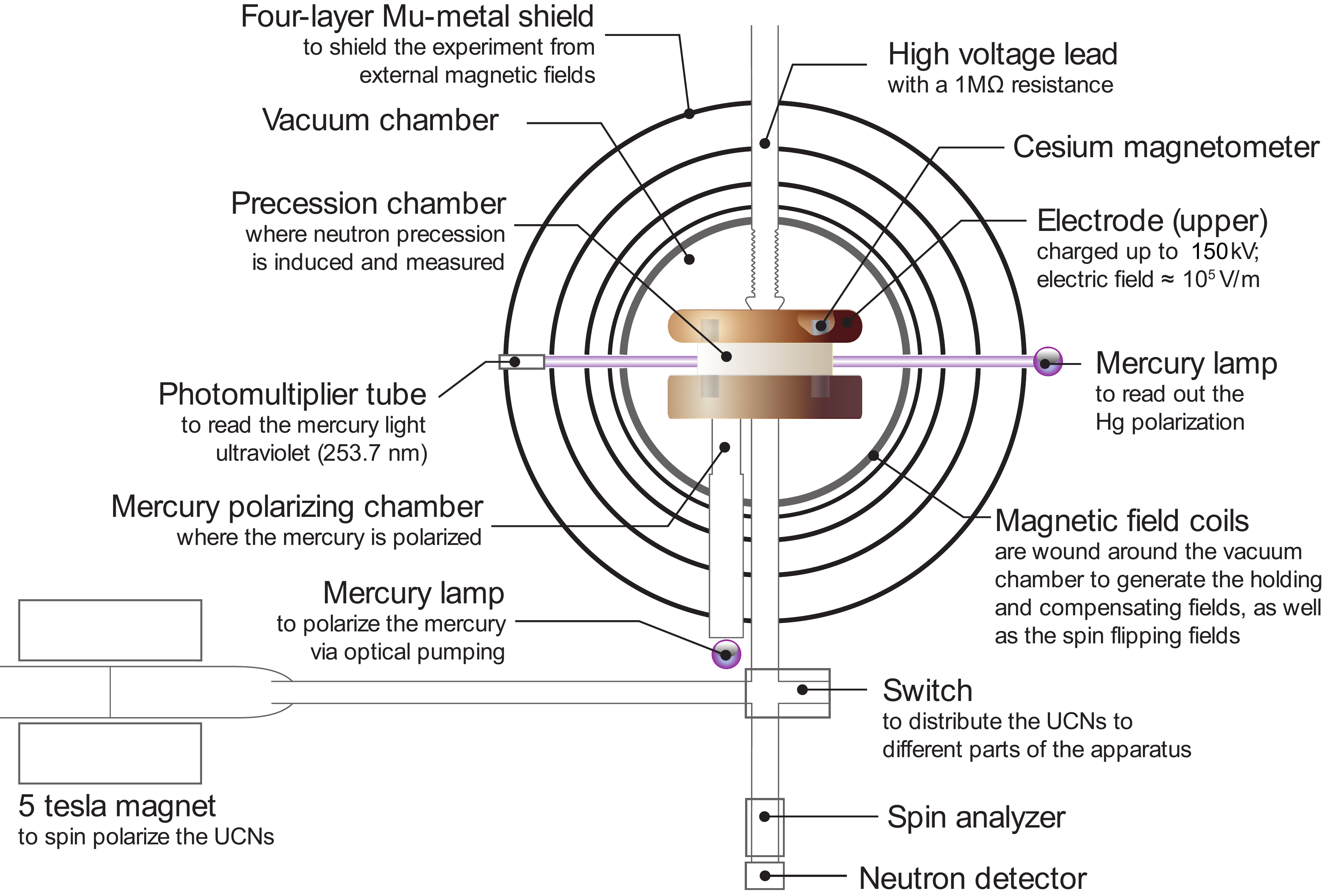}
\caption{Sketch of the present nEDM apparatus set up a the Paul
Scherrer Institut.} \label{fig:spectrometer}
\end{figure*}

Figure\,\ref{fig:spectrometer} is a sketch of the RAL/Sussex/ILL
apparatus installed at the new UCN source at PSI. Neutrons come in
from the left and are polarized to $100$\% by a $5$\,T
superconducting solenoid. They are then guided through a UCN switch
upwards into the precession chamber. Once an equilibrium UCN density
has built up the UCN shutter on the bottom electrode is closed. Now
a shutter is opened for $2$\,s such that polarized \magHg{} atoms
can also fill the chamber. Neutrons and \magHg{} atoms are polarized
parallel or anti parallel to the main vertical magnetic field $B_0
=1\,\mathrm{\mu T}$. First, the \magHg{} atoms are spin flipped into
the plain perpendicular to $B_0$, by applying a $\sim 8$\,Hz
$\pi/2$-pulse, then the neutrons ($30$\,Hz). Now both spin-$1/2$
species precess with their Larmor frequency within the precession
chamber. After a time~$T$ a second 30\,Hz pulse, in-phase with the
first, is applied before the UCN shutter is opened. Meanwhile the
switch has changed to a position connecting directly the precession
chamber with the UCN detector. An iron foil in front of the detector
together with an adiabatic spin flipping rf~coil located further up
stream allows to measure both spin states sequentially at the end of
a cycle. Such a sequence of two neutron $\pi/2$-pulses with the free
precession time $T$ in between is known as Ramsey's method of
separated oscillating fields and correlate the polarization measured
with the free precession frequency and the frequency of the applied
fields.

A circularly polarized light beam from a $ ^{204}$Hg~discharge lamp
($\lambda = 254$\,nm) is used to polarize a vapor of \magHg{} atoms
by optical pumping in a cell adjacent to the main precession
chamber. The \magHg{} atoms dissociate from a solid HgO source
heated to $200\,\mathrm{^{\circ}C}$, and continuously fill the cell
for the next Ramsey cycle.

A second circularly polarized light beam traverses the precession
chamber and is used to read out the precessing frequency of the
\magHg{} atoms. Its intensity is measured by a solar-blind photo
multiplier~(PM). After application of the $\pi/2$-pulse the
precessing \magHg{} atoms modulate the light intensity at the Larmor
frequency of the precessing atoms proportional to the average
magnetic field inside the chamber.

Twelve Cs-OPM, four high voltage compatible ones on the charged
electrode the others beneath the ground electrode, measure the
magnetic field throughout the cycle. Although the intrinsic
sensitivity of the magnetometers is of the order of $10\,\fTHz$,
technical noise and the magnetic environment presently limit the
sensitivity to $\sim1\,\pTHz$.


\section{Systematic effects}
Decreasing the statistical uncertainty by a factor five also
requires to decrease uncertainties from systematic effects
accordingly. This can be achieved by realistic technical
modifications and upgrades\,\cite{Altarev2009}. In the last two
years we were trying to asses most of these systematic errors by
measurements without neutrons. In the remainder of these proceedings
we will focus in more detail on the uncompensated field drift and
give just a brief status of systematic effects directly correlated
with changing the polarity of the upper electrode:


{\it Leakage currents} which might flow from the charged top
electrode to the ground electrode are monitored with a custom made
A/V amplifier, protected by an in-house designed protection circuit.
This device allows us to measure and monitor leakage currents with a
resolution of some $10\,\mathrm{pA/\sqrt{Hz}}$. It was used during
an intensive test campaign with voltages of up to $\pm100$\,kV and
shows that typical leakage currents are below 100\,pA, resulting in
a sensitivity of $0.01\times10^{-27}\,\ecm$

{\it Hg direct light shift} is a shift of the measured \magHg{}
precession frequency proportional to the intensity of the UV light
beam traversing the precession chamber. The sign of the shift
depends on the relative direction of the electric and magnetic field
and therefore mimics the signal of a nEDM. Using data from spring
2011 we could not observe such a dependency and hence estimate a
sensitivity of $0.04\times10^{-27}\,\ecm$.

{\it Hg light shift}, a second type of mercury frequency shift, can
be observed, when the read-out light is not exactly perpendicular to
the applied $B_0$ field. A new beam optics now focus the beam onto
the PM. From mechanical measurements one can estimate that the light
beam axis diverges at an angle less than 8\,mrad from a plain
parallel to the electrodes, resulting in the projected sensitivity
of $0.4\times10^{-27}\,\ecm$.

\section{Measurement of the uncompensated field drift}
`Uncompensated field drift' refers to systematic effects which arise
from higher order magnetic fields the mercury co-magnetometer can
not completely compensate for. These drifts can arise from charging
currents magnetizing parts of the apparatus while changing the
polarity of the high voltage. This creates a change in the local
magnetization and hence a change of the vertical magnetic field
gradient $\Delta G = (\tfrac{\partial B}{\partial
z}_{\parafield}-\tfrac{\partial B}{\partial z}_{\aparafield})$
dependent on the polarity of the high voltage. This effect yields a
false nEDM signal of:

\begin{equation}
    d_{\sub{f}} =
    \left|\frac{\gamma_{\sub{n}}}{\gamma_{\sub{Hg}}}\right|\frac{\hbar\omega_{\sub{Hg}}\Delta
    h}{4EB_0}\Delta G.
\end{equation}

\begin{figure*}[t]
\centering
\includegraphics[width=135mm]{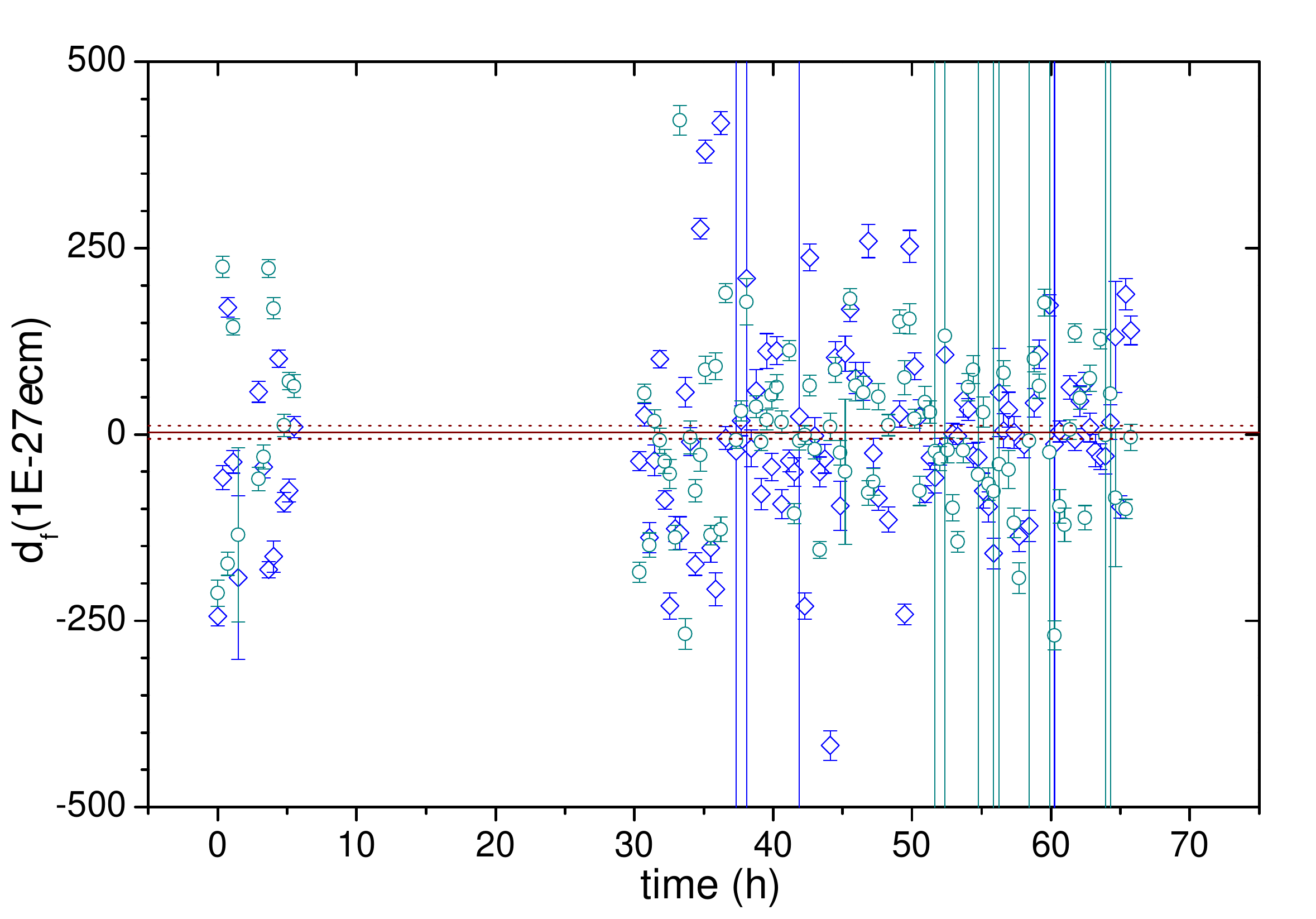}
\caption{False nEDMs from uncompensated field drifts, measured with
two pairs of Cs-OPM
(Cs1-Cs2:\,{\color{Blue}$\Diamond$},Cs3-Cs4:\,{\color{JungleGreen}$\ocircle$}).
The horizontal line is the weighted mean of the combined data sets.
The dotted line indicate the weighted standard deviation.}
\label{fig:Uncompensated_field_data}
\end{figure*}

In autumn 2010, we searched directly for a change in the vertical
gradient of the magnetic field with four Cs-OPM. Always a pair
separated by $\sim 20$\,cm in height, one on the top electrode, the
other below the bottom electrode, was taken as gradiometer (Cs1-Cs2,
Cs3-Cs4). The magnetic background field was set to $B_0\approx
1\,\mathrm{\mu T}$. The polarity of the electric field ($E=\pm
100$\,kV) was changed every 350\,s with the system's maximum ramping
speed of 1\,kV/s. In Fig.\,\ref{fig:Uncompensated_field_data} data
sets for a false $d_{\sub{f}}$ for both Cs-OPM pairs, taken during
$\sim 70$\,h, are shown.

A linear correlation analysis shows that both data sets can be
combined. This first test gives a preliminary value of
$d_{\sub{f}}=2.9\pm8.6 \times 10^{-27}\,\ecm$. Our goal is to
improve this value by a factor $\sim\,10$ which seems possible in
one week of measurements by optimizing the measurement sequence to
$260$\,s (instead of $1100$\,s), increasing the maximum voltage to
$\sim 120$\,kV, and using four independent pairs of Cs-OPM. At the
same time we will use an active compensation system around our
apparatus to better compensate external magnetic field changes.

\section{Conclusion and Outlook}
While the new PSI UCN source is ramping up our collaboration is
prepared to take first nEDM data in autumn 2011. During the winter
shutdown of the accelerator we plan to continue our study of crucial
systematic effects. By the end of 2013 we expect to have reached our
required sensitivity through 400 nights of good quality data. This
then would result in the case of a null result in a new upper limit
of $\nedm < 5\times 10^{-27}\ecm$\,(95 \,C.L.). In parallel we are
constructing the next generation apparatus.

\begin{acknowledgments}
We are grateful for the technical support throughout the
collaboration and the PSI workshops. We would like to thank the
groups from Technical University Munich and DFG cluster of
excellence ``Origin and Structure of the Universe'' for their
support and many discussions.

This work was supported by the Swiss National Science Foundation by
grants No. 200020-130480 and 200021-126562
\end{acknowledgments}

\bigskip

\end{document}